\documentstyle[12pt,epsf]{article}

\def\avbr {B_{avg}}

\begin{document}

\begin{flushright}
FERMILAB-PUB-95/179-E\\
LNS-95-157\\
July 26, 1995\\
\end{flushright}

\vskip1cm

\begin{center}
\LARGE{\bf Extracting $\alpha$ from the
CP Asymmetry in $B^0 \to \pi^+ \pi^-$ Decays}\\
\vskip 1 cm
\large{F. DeJongh}\\
\large{\it Fermilab}\\
\vskip 0.5 cm
\large{P. Sphicas}\\
\large{\it Massachusetts Institute of Technology}\\
\end{center}
\vskip1cm

\begin{abstract}
The extraction of the CKM angle $\alpha$ from the asymmetry
in $B^0 \to \pi^+\pi^-$ vs ${\bar B^0} \to \pi^+\pi^-$
suffers from a currently unknown penguin contribution.
Experimentally, one can determine the magnitude and phase
of the CP asymmetry from a time-dependent analysis of
tagged events, and
the average rate for $B^0$ and $\bar{B}^0$ decays
to $\pi^+\pi^-$ from untagged events.
These measurements, together with the magnitudes and relative phase
of the tree and penguin diagrams,
can in principle completely determine $\alpha$,
free of discrete ambiguities.
We perform an error analysis on $\alpha$ given assumptions
on the values and uncertainties of both the
measurements and theoretical inputs.
\end{abstract}

\newpage

\section{Introduction}
\label{section_intro}

The unitarity of the CKM matrix corresponds to six independent
conditions between

\vskip 5 pt
\parindent 0pc \leftmargini 0em \leftmargin\leftmargini
\parbox{3.25 in}
{the CKM matrix elements\cite{Bdecays_book}.
Geometrically, these conditions can be visualized as
triangles formed by the appropriate CKM matrix element combinations.  The
condition on the $d$ and $b$ rows, $\sum_{q=u,c,t} V_{qd}V_{qb}^* = 0$,
results in the CKM triangle shown in Fig.~1.  If CP violation occurs via
the CKM matrix, this triangle will have non-zero area.
The three angles in this triangle, known as $\alpha$, $\beta$ and $\gamma$,
can be extracted via CP asymmetries:
the angle $\beta$ from the decay $B^0 \to J/\psi K_s$\cite{CP_beta},
the angle $\alpha$ from the decay $B^0 \to \pi^+\pi^-$\cite{Gronau-London}
and the angle $\gamma$ from the decays
$B_s \to D_s K$\cite{CP_gamma_1} and
$B^\pm \to D^0 K^\pm$\cite{CP_gamma_2}.
}
\parbox{2.75 in}
{\begin{picture}(195,140)(0,0)
\put(20,10){\line(1,0){170}}
\put(190,10){\line(-3,4){90}}
\put(100,130){\line(-2,-3){80}}
\put(85,-10){$V_{cd}V_{cb}^*$}
\put(145,80){$V_{td}V_{tb}^*$}
\put(30,80){$V_{ud}V_{ub}^*$}
\put(40,20){$\gamma$}
\put(160,20){$\beta$}
\put(95,110){$\alpha$}
\end{picture}
\center{Figure 1: The $d,b$ CKM triangle}
}

\parindent 1 cm
\leftmargini 2.5em

The amplitude for $B \to \pi\pi$ is dominated
by the tree process $b \to uW, W \to d\bar{u}$, which
has a weak phase $\gamma$.  If this were the only contribution,
the CP asymmetry in $B^0 \to \pi^+\pi^-$ would cleanly measure
$\sin(2\beta + 2\gamma) = -\sin(2\alpha)$.
However, there may be significant contributions from
the penguin process $b \to dg$.  This process has a weak phase
$-\beta$ and therefore results in a distortion of the CP asymmetry.
Any interpretation of the CP asymmetry in $B^0 \to \pi^+\pi^-$
must therefore account for the possibility of a penguin contribution.

Gronau and London\cite{Gronau-London} have shown that
the penguin contributions can be isolated by applying an
isospin analysis to the decays
$B^0\to \pi^+\pi^-$, $B^+ \to \pi^+\pi^0$ and $B^0\to \pi^0\pi^0$.
Aleksan, Gaidot, and Vaisseur\cite{GLerror} have estimated that
this analysis typically results in a 60\% increase in the uncertainty
on $\sin(2\alpha)$ relative to the ideal case where only the tree
diagram need be considered.
While feasible for an experiment at an $e^+e^-$ collider,
this analysis is of no use to an experiment
at a hadron collider, given that it is very unlikely that the mode
$B^0\to \pi^0\pi^0$ will ever be reconstructed in such an environment.

Silva and Wolfenstein have shown\cite{Silva} that
$\alpha$ can be determined from the CP asymmetry in
$B^0 \to \pi^+\pi^-$, the relative rates for
$B^0 \to \pi^+\pi^-$ and $B^0 \to K^+\pi^-$,
and assuming SU(3) symmetry and factorization.
Some complications are the possibility of final state phase
shifts, and electroweak penguins
that invalidate the SU(3) correspondence\cite{DH1}.

We present herein an analysis of the expected uncertainty on $\alpha$,
and the number of discrete solutions, given a measurement of
the time-dependent asymmetry between
$B^0 \to \pi^+\pi^-$ and ${\bar B^0} \to \pi^+\pi^-$,
a measurement of the average branching ratio
for $B^0$ and ${\bar B^0}$ decays to $\pi^+\pi^-$,
and constraints on the magnitudes and relative phase
of the tree and penguin diagrams.

\section{CP violation in $B^0 \to \pi^+\pi^-$}
\label{section_theory}

The mathematical expression
of the CP asymmetry in the decay $B^0 \to \pi^+\pi^-$
can be found in numerous places in the literature.  Here we follow
the exposition of Gronau\cite{penguins2}.

In general, the two physical states $B_L$ and $B_H$ are given in
terms of the strong eigenstates $B^0$ and $\bar B^0$ via
\begin{eqnarray}
|B_L\rangle = p |B^0\rangle + q |{\bar B^0}\rangle \\
|B_H\rangle = p |B^0\rangle - q |{\bar B^0}\rangle
\end{eqnarray}
If two amplitudes (e.g. tree-level and penguin) contribute to the
decay $B^0 \to f$, then the decay amplitudes of $B^0$ and $\bar B^0$
to a CP eigenstate $f$ are given by
\begin{eqnarray}
\label{eqamp}
A_f & = & A(B^0 \to f) = A_T e^{i\phi_T} + A_P e^{i\phi_P} \\
{\bar A_f} & = & A({\bar B^0} \to f) = A_T e^{-i\phi_T} + A_P e^{-i\phi_P}
\end{eqnarray}
where each term in the above expression corresponds to a process.  The
amplitudes $A_i$ are complex and contain hadronic final-state-interaction
phases.
The time-evolution of states initially pure in $B^0$ and $\bar B^0$ are
then given by
\begin{eqnarray}
\Gamma(B^0 \to f) & = & |A_f|^2 e^{-t}
\left [
|\lambda|^2 \sin ^2 (xt/2) + \cos ^2 (xt/2) - Im{\lambda}\sin xt
\right ] \\
\Gamma({\bar B^0} \to f) & = & |A_f|^2 e^{-t}
\left [
\sin ^2 (xt/2) + |\lambda|^2 \cos ^2 (xt/2) + Im{\lambda} \sin xt
\right ]
\label{eqgamma}
\end{eqnarray}
where
\begin{equation}
\lambda = \frac{q}{p} \frac{\bar A_f}{A_f}
\end{equation}
The time-dependent asymmetry, $a(t)$, is thus
\begin{eqnarray}
\label{eq1}
a(t) = \frac{\Gamma(B^0 \to f) - \Gamma({\bar B^0} \to f)}
	{\Gamma(B^0 \to f) + \Gamma({\bar B^0} \to f)}
= \frac{ (1 - |\lambda|^2) \cos xt - 2Im\lambda\sin xt}{1+|\lambda|^2}
\end{eqnarray}
For the case $f=\pi^+\pi^-$, $q/p = e^{-2i\beta}$ and
$\phi_T = \gamma$, and in the approximation of neglecting the penguin
contribution, i.e. $A_P = 0$, $\lambda$ is a pure phase which
results in $Im\lambda = -\sin(2\beta + 2\gamma) = \sin(2\alpha)$,
assuming the unitarity of the CKM matrix.
In this case, the amplitude of the asymmetry directly yields
a clean extraction of the angle $2\alpha$ -- but with a discrete ambiguity.

In this same decay mode, however, the penguin, assumed to be dominated
by the top quark loop, has a CKM phase given by
$\phi_P=-\beta$ and therefore the extraction of $2(\beta + \gamma)$
is not clean.  Inspection of equation~\ref{eq1} shows that the
effect of the penguin contribution is the addition of an additional
sinusoidal modulation in the time-dependent
asymmetry, the additional factor $(1 - |\lambda|^2) \cos xt$.  The overall
asymmetry can then be written as
\begin{eqnarray}
a(t) = A \sin(xt + \phi)
\end{eqnarray}
where
\begin{eqnarray}
A & = & \frac{\sqrt{(1 - |\lambda|^2)^2 + 4Im^2\lambda}}{1+|\lambda|^2}
 \times {\rm sign}(-Im\lambda) \\
\tan\phi & = & \frac{1 - |\lambda|^2}{2 Im\lambda},
 -\frac{\pi}{2}<\phi<\frac{\pi}{2}
\end{eqnarray}
This convention reduces smoothly to the standard expression for the
no-penguin case.
We also exploit the dependence of the average branching ratio for
$B\to\pi^+\pi^-$, $\avbr$ on the angle $\alpha$:
\begin{eqnarray}
\avbr \propto {\Gamma(B^0 \to \pi^+\pi^-)+
		\Gamma({\bar B^0} \to \pi^+\pi^-)} =
		|A_f|^2 \left [1 + |\lambda|^2 \right ]
\end{eqnarray}

In what follows we will refer to the strength of the penguin contribution,
$A_P$, relative to the tree-level, $A_T$.
We thus introduce the ratio of the amplitudes
$f=A_P/A_T=|f|e^{i\delta}$, where $\delta$ is the strong phase difference
between the amplitudes, and obtain for $\lambda$:
\begin{eqnarray}
\lambda = e^{2i\alpha}\frac{1-fe^{-i\alpha}}
{1-fe^{i\alpha}}
\label{eqlambda}
\end{eqnarray}

\begin{figure}[hp]
\epsfysize=6.0in
\centerline{
\epsffile{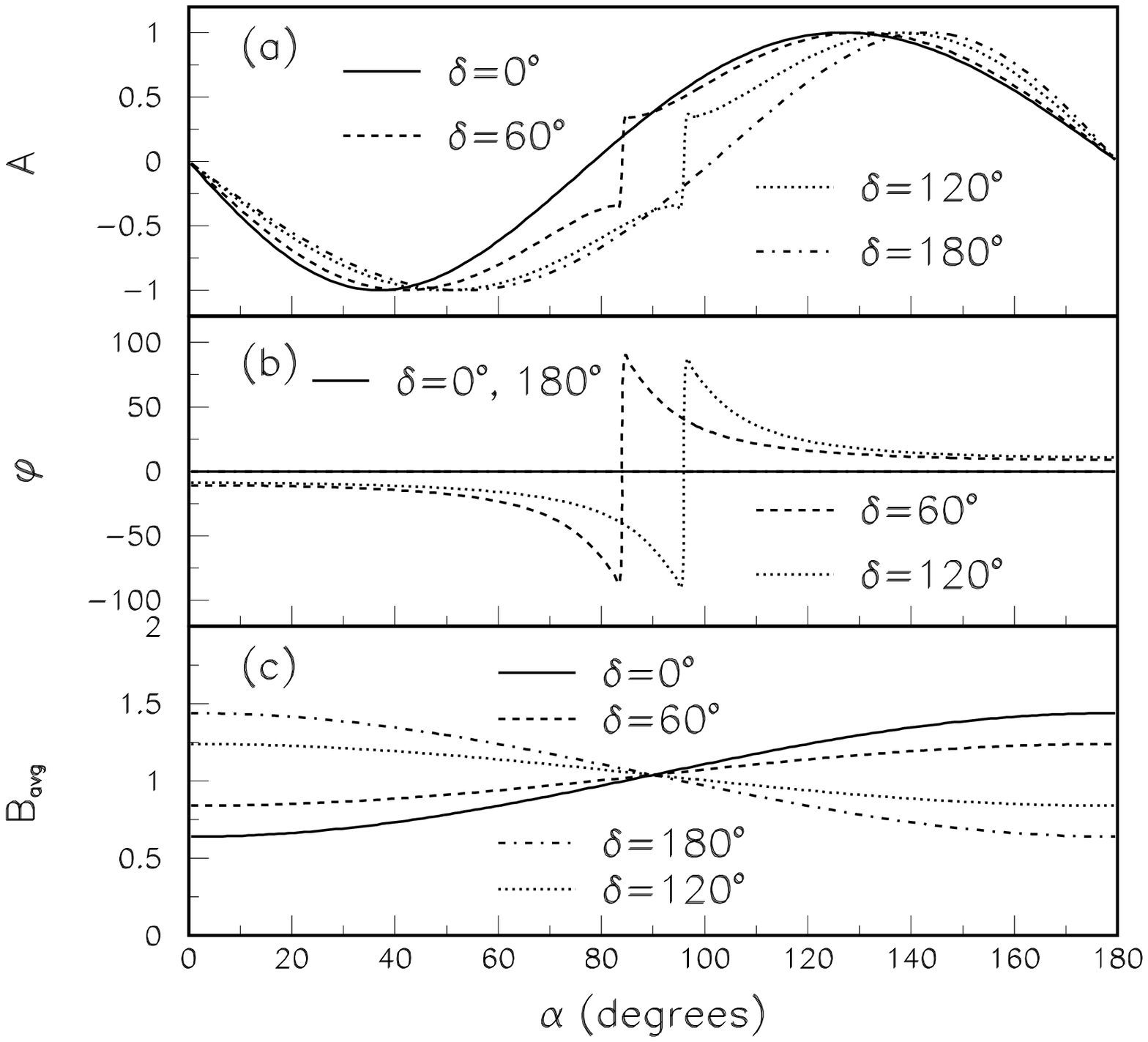}}
\caption{The three experimental quantities, the asymmetry, $A$,
the phase, $\phi$, and the average branching ratio, $\avbr$
(normalized to the ``no penguin'' case), as function
of the angle $\alpha$.  The relative size of the penguin contribution
is $20\%$.  The various curves correspond to different values of the
strong phase difference $\delta$.}
\label{theory_2}
\end{figure}

Experimentally, we have three observables: the magnitude of the
asymmetry, $A$, the phase of the asymmetry at $t=0$, $\phi$, and the
average branching ratio, $\avbr$.  All of these are affected by both the
tree-level and penguin diagrams.
The dependence of the three observables, $A, \phi, \avbr$,
on the angle $\alpha$ is shown in figure~\ref{theory_2},
for various values of $\delta$.  We have
taken $|f|=0.2$ for these plots.  We see that the asymmetry
in the presence of a penguin contribution is no longer symmetric
around $\alpha=90^o$.  Also, at $\alpha=90^\circ$, the average branching
ratio is no longer equal to the tree-level one, but it
is increased by a factor $1+|f|^2$, i.e. 1.04 in this example.
Note that the phase, $\phi$, vanishes for $\delta=0,180^o$.

In the absence of penguins, there is a four-fold ambiguity in the
determination of $\alpha$ from the CP asymmetry:  The asymmetry is
identical for the case $\alpha \to \pi/2 - \alpha$, and for
$\alpha \to \alpha - \pi$.  A most interesting
feature of the plots in figure~\ref{theory_2} is the behavior of the
branching ratio between $0$ and $\pi/2$: the curves change monotonically
and thus lift the ambiguity between
$\alpha$ and $\pi/2 - \alpha$.
Also, the curve for $A$ is antisymmetric around $\alpha = 0$, while the
curves for $\phi$ and $\avbr$ are symmetric around $\alpha = 0$.  Thus,
the ambiguity between $\alpha$ and $\alpha - \pi$ is also lifted:  there
is only a single discrete case, $A=0$ and $\alpha \to -\alpha$, where two
values of $\alpha$ are a solution given $A$, $\phi$, and $\avbr$.
In summary, in the presence of penguins, the four-fold ambiguity is
in principle completely lifted except for the discrete case $A=0$
where it becomes a two-fold ambiguity, $\alpha \to -\alpha$.

In the next section we estimate the expected error on $\phi$ from
fitting the above asymmetry as a function of the statistics.
Most proposals for experiments at hadron colliders involve a
$\pi^+ \pi^-$ trigger that imposes
(usually indirectly, via impact parameter requirements)
an effective cut $t>T$ on the  lifetime of the $B^0$ decays
reconstructed.  We thus compute the error on the observables as
a function of the effective cut value, $T$.

\section{Fitting the data for $A$ and $\phi$}
\label{section_fitting}

The numbers of $B^0$ ($N_+(t)$) and $\bar B^0$ ($N_-(t)$)
at time $t$ can be written as
\begin{eqnarray}
\label{eqn}
N_\pm(t) = \frac{N}{2} e^{-t} \left [ 1 \pm A \sin ( xt +\phi) \right ]
\end{eqnarray}
where we used equation~\ref{eqgamma} and integrated over all time
to express $|A|^2$ in terms of the total number of $B^0$ and $\bar B^0$
mesons, $N$.
We are interested in estimating the error on the quantities $A$ and $\phi$
resulting from a fit to the data by the above form.  The probability that
a set of $B^0$ and $\bar B^0$ mesons (initially pure) will
decay at times $t_i$ and $t_j$ respectively is given by
\begin{eqnarray}
\label{likelihood}
{\cal L} = \prod_{i} e^{-t_i} \left [ 1 + A \sin ( xt_i +\phi) \right ]
	   \prod_{j} e^{-t_j} \left [ 1 - A \sin ( xt_j +\phi) \right ]
\end{eqnarray}

\begin{figure}[hp]
\epsfysize=5in
\centerline{
\epsffile{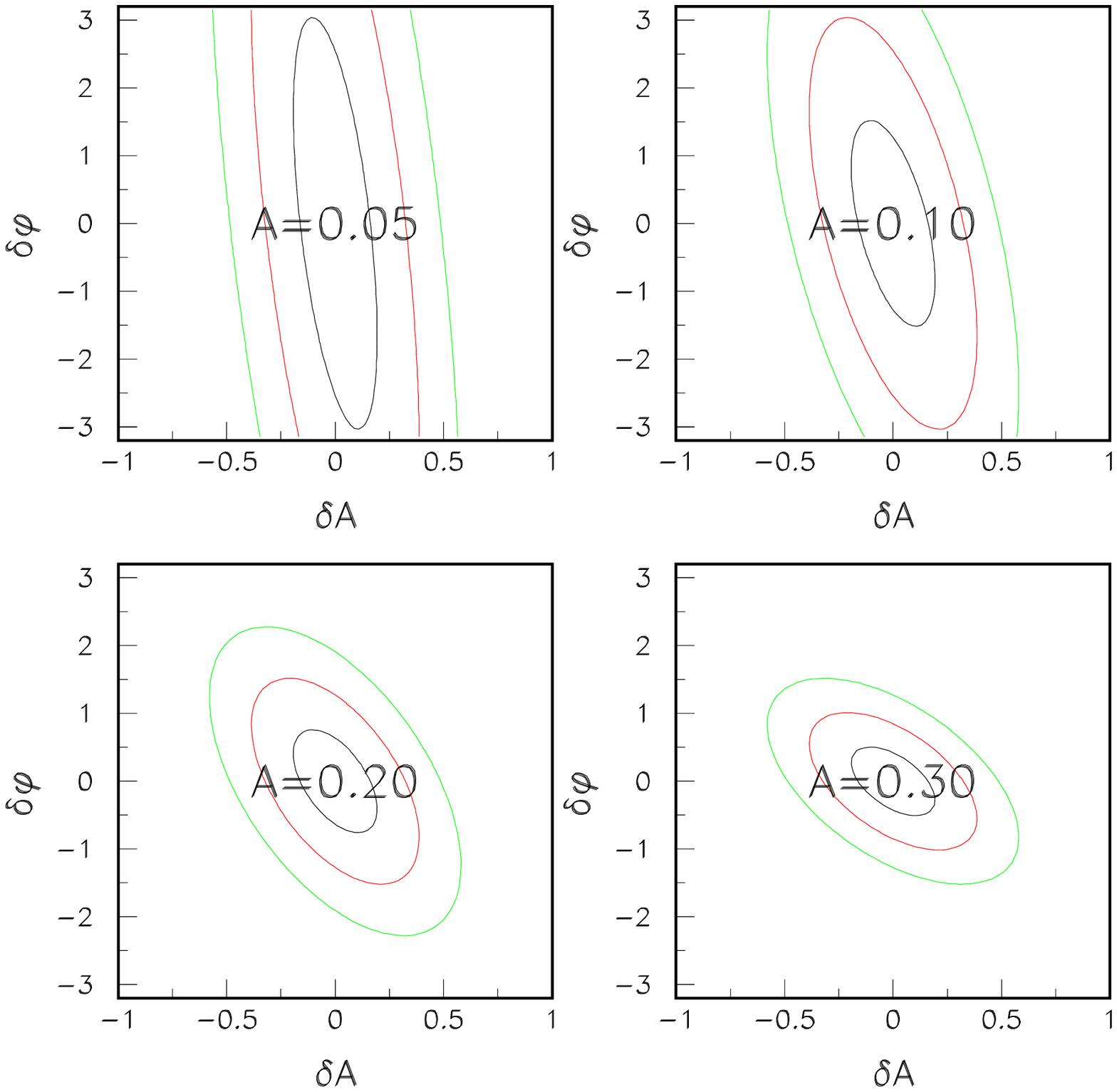}}
\vspace{-1cm}
\caption{Contours of equal probability in the $\delta A-\delta\phi$ plane}
\label{ellipse}
\end{figure}

Equation~\ref{likelihood} is then the expression for the likelihood for
this set of events.  The inverse of the covariance matrix for
the variables $A$ and $\phi$, $G = V^{-1}$,
is given by
\begin{eqnarray}
\label{eqnerrors}
G_{AA} = -\frac{\partial^2 \ln {\cal L}}{\partial A^2}
\ \ \ \
G_{\phi\phi} = -\frac{\partial^2 \ln {\cal L}}{\partial \phi^2}
\ \ \ \
G_{A\phi} = -\frac{\partial^2 \ln {\cal L}}{\partial A \partial\phi}
\end{eqnarray}
The covariance matrix is computed in appendix A.

%

The one sigma contour in the $A-\phi$ plane is given by the ellipse
equation:
\begin{eqnarray}
\label{ellipseq}
\frac{1}{1-\rho^2} \left[ \frac{\delta A^2}{\sigma^2_A}
	- \frac{2\rho\delta A\delta\phi}{\sigma_A\sigma_\phi}
	+ \frac{\delta\phi^2}{\sigma^2_\phi} \right] = 1
\end{eqnarray}
where $\delta A= A-{\bar A}$, $\delta\phi = \phi - {\bar \phi}$ and ${\bar A}$,
${\bar \phi}$ are the ``true'' values, and $\rho$ is the correlation
coefficient.
This error ellipse is shown in figure~\ref{ellipse}
for four different values of the asymmetry $A$, and $\phi=0.1$.
It can be seen that the error on
$\phi$ decreases as the asymmetry gets larger.


\section{Extraction of $\alpha$ given $A$, $\phi$, and $\avbr$}
\label{section_extraction}

As discussed in Section~\ref{section_theory}, there are three observables
related to the CP asymmetry in $B \rightarrow \pi\pi$.
The uncertainties on the measurements of $A$ and $\phi$ were discussed
in Section~\ref{section_fitting}.  The third observable,
$\avbr$, can be measured from untagged events, and will therefore
have a very low statistical error.  The extent to which
systematic uncertainties can be controlled will therefore be a
crucial consideration.  While absolute
branching ratios are very difficult to obtain, it will be sufficient
to measure the branching ratio relative to the process
$B^0 \to \ell^+ \pi^- \nu$.

There are five unknowns that influence the values of the 3 observables:

\begin{itemize}
\item $\alpha$, the weak phase we are trying to extract from these
      measurements.
\item $A_T$, the amplitude of the tree diagram.  Assuming factorization,
      the amplitudes for $B^0 \to \ell^+ \pi^- \nu$ and
      $B^0 \to \pi^- \pi^+$ are proportional to a common form-factor,
      spanning a range of $q^2$ for the first case, and evaluated at
      $q^2 = m_\pi^2$ for the latter case\cite{BSW}.
      Ref.~\cite{Kamal} points out that color-allowed $B$ decays are well
      described by factorized amplitudes, but find that it is
      necessary to add non-factorized amplitudes to describe
      color-suppressed $B$ decays.
      Since the decay $B^0 \to \pi^- \pi^+$ is color allowed, we assume
      that the decay $B^0 \to \ell^+ \pi^- \nu$ will be observed in
      conjunction with $B^0 \to \pi^- \pi^+$ and used to predict $A_T$.
\item $A_P$, the amplitude of the penguin diagram.  This amplitude can
      be estimated from measurements of the decay $B^0 \to K^- \pi^+$,
      applying SU(3) corrections, and scaling by
      $|V_{td} / V_{ts}|$\cite{Silva}.
      Some complications are that this decay may in turn have a contribution
      from tree diagrams, and furthermore, there may be electroweak
      contributions that invalidate the SU(3) correspondence\cite{DH1}.
      The decay mode $B_s \to \phi \rho^0$ may possibly be used
      to check our
      understanding of these effects\cite{DH2}.
      We assume that the relative size of the signals for
      $B^0 \to K^- \pi^+$ and $B^0 \to \pi^- \pi^+$ will
      determine $A_P$, although with less precision than for $A_T$.
\item The weak phase of the penguin amplitude.  The top quark
      dominates in the loop, therefore this phase will be $-\beta$ to a
      good approximation\cite{penguins1}.  As shown in
      equation~\ref{eqlambda}, in this case we are not sensitive to the
      value of $\beta$.
\item The strong phase difference, $\delta$, between the tree and
      penguin diagrams.  There is a perturbative phase difference of
      order $10^\circ$\cite{DH1},
      and there are also nonperturbative effects from
      hadronization that are expected to be small but are incalculable.
      The phase shift between the I=0 and I=2 amplitudes can be obtained
      from a measurement of the branching ratios of
      $B \to \pi^+\pi^-$, $B \to \pi^+\pi^0$, and $B \to \pi^0\pi^0$.
      This check can be done at a symmetric $e^+e^-$ collider, and does
      not require flavor tagging or time-ordering.  The question then
      becomes:  Is there a difference in the hadronization for a penguin
      diagram and I=0 tree diagram?
      The extent to which these phase shifts can be constrained helps
      constrain our solution for $\alpha$.
\end{itemize}

Given an assumption for the central values of these parameters, we can
calculate the values of $A$, $\phi$, and $\avbr$.  Given an assumption for the
effective number of tagged events, $N_{tag}$\cite{dilution},
we can calculate the error matrix for $A$ and $\phi$.
We will make assumptions on the uncertainties
on $A_T$, $A_P$, $\delta$, and $\avbr$, parametrized as Gaussians with widths
$\sigma(A_T)$, $\sigma(A_P)$, $\sigma(\delta)$, and $\sigma(\avbr)$.
With these assumptions, we can form a $\chi^2$.  The minimization
program MINUIT is used to minimize this $\chi^2$,
return the input value of $\alpha$, and estimate the expected uncertainty
on a measurement of $\alpha$.

\begin{figure}[hp]
\epsfysize=6in
\centerline{
\epsffile{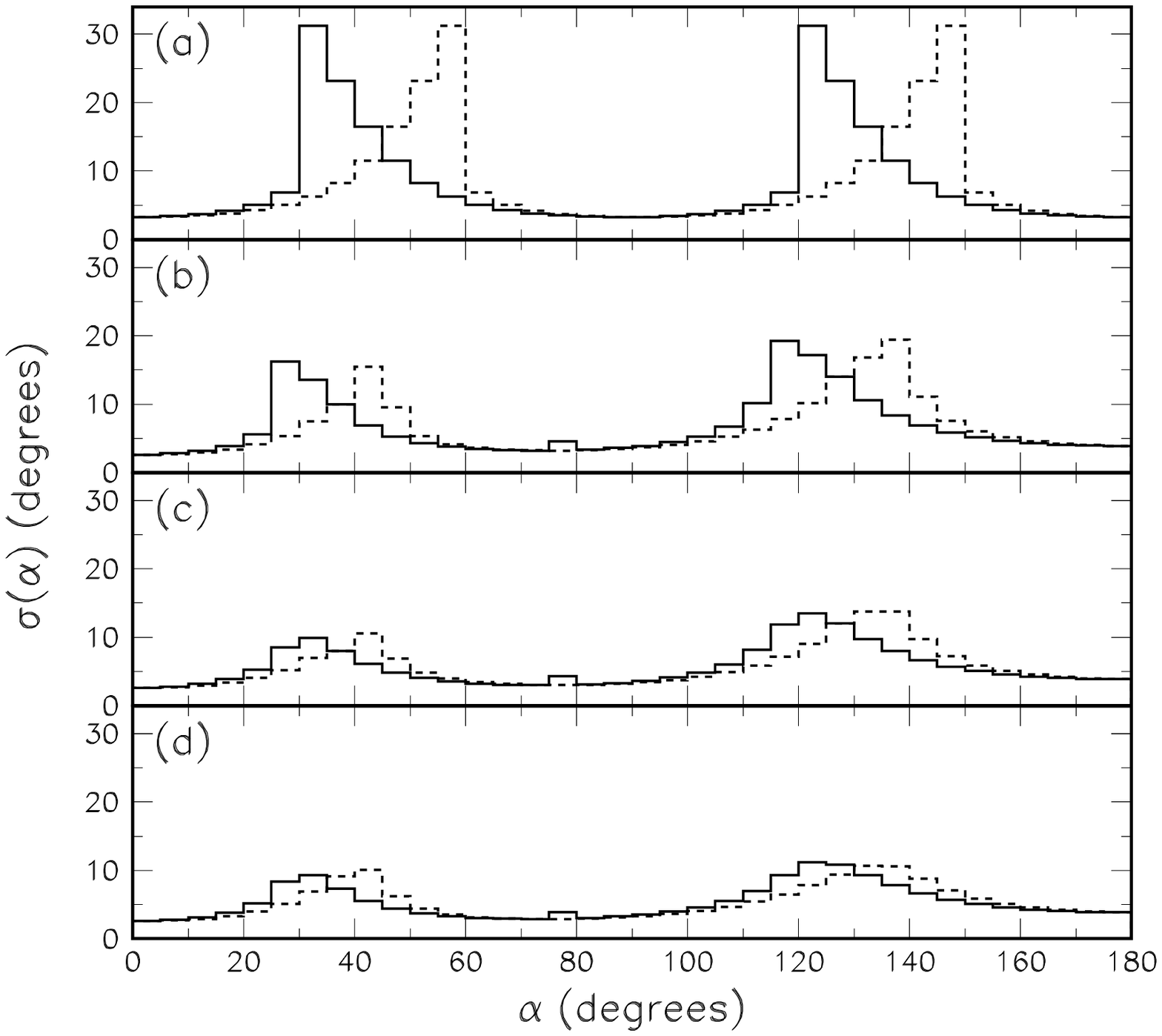}}
\caption{One $\sigma$ errors on $\alpha$ as a function of the input value
of $\alpha$.  The solid line shows the positive errors, and the dashed line
the negative errors: \hfil\break
(a)  $A_P = 0$, and fixed to 0 in the fits. \hfil\break
The following assume $A_P = 0.2$: \hfil\break
(b)  $\sigma(A_P)/A_P = 0.1$,
$\sigma(A_T)/A_T = 0.03$, $\sigma(\delta) = 20^\circ$, and
$\sigma(\avbr)/\avbr = 0.06.$ \hfil\break
(c)  $A_P$, $A_T$, and $\delta$
held fixed in the fits, and $\sigma(\avbr)/\avbr = 0.06.$ \hfil\break
(d)  $\sigma(A_P)/A_P = 0.1$,
$\sigma(A_T)/A_T = 0.015$, $\sigma(\delta) = 20^\circ$, and
$\sigma(\avbr)/\avbr = 0.03.$
}
\label{onesigma}
\end{figure}

Unless specfied otherwise, we use the following as default values of the
parameters:
\begin{itemize}
\item $N_{tag} = 100$, starting at $c\tau$ = 1.6 lifetimes\cite{Lewis}.
\item $A_T = 1.0$
\item $A_P = 0.2$
\item $\delta = 0.0$
\end{itemize}

\begin{figure}[hp]
\epsfysize=6in
\centerline{
\epsffile{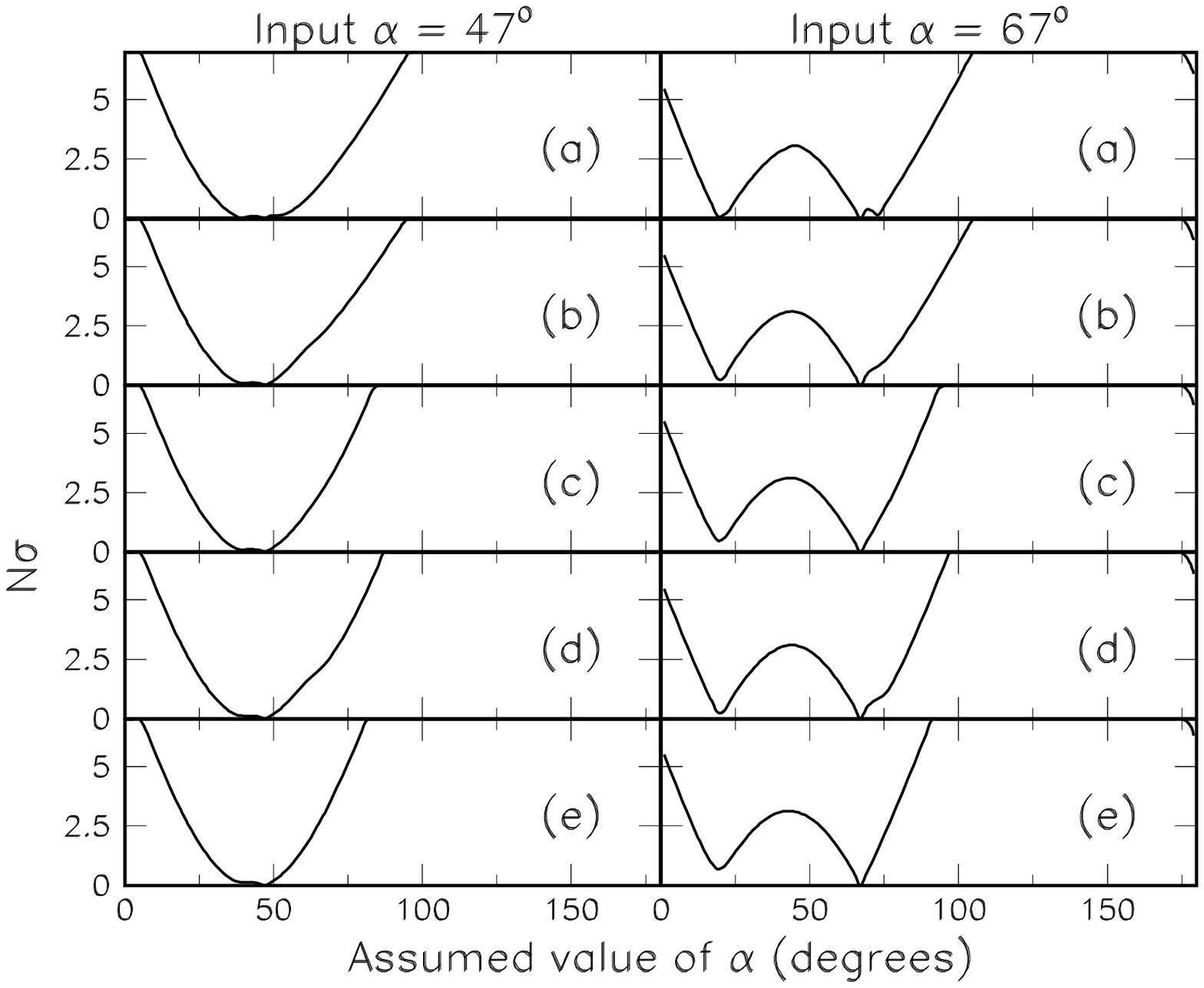}}
\caption{$N\sigma$ as a function of $\alpha$, for an input value of
$\alpha = 47^\circ$ on the left and $\alpha = 67^\circ$ on the
right.  The input value for $A_P/A_T$ is 0.05.
\hfil\break
For the following, we assume $\sigma(\avbr)/\avbr = 0.06$: \hfil\break
(a)  $\sigma(A_P)/A_P = 1.0$,
$\sigma(A_T)/A_T = 1.0$, no constraint on $\delta$. \hfil\break
(b)  $\sigma(A_P)/A_P = 1.0$,
$\sigma(A_T)/A_T = 0.03$, no constraint on $\delta$. \hfil\break
(c)  $\sigma(A_P)/A_P = 1.0$,
$\sigma(A_T)/A_T = 0.03$, $\sigma(\delta) = 20^\circ$. \hfil\break
(d)  $\sigma(A_P)/A_P = 0.1$,
$\sigma(A_T)/A_T = 0.03$, no constraint on $\delta$. \hfil\break
(e)  $\sigma(A_P)/A_P = 0.1$,
$\sigma(A_T)/A_T = 0.03$, $\sigma(\delta) = 20^\circ$.
}
\label{ap05}
\end{figure}

In Fig.~\ref{onesigma} we show the expected $1\sigma$ errors as a function
of $\alpha$ for various conditions.  In Fig.~\ref{onesigma}a, we show the
errors for the case where $A_P = 0.0$, and has been constrained to zero
in the fit.
We see that the errors are largest for $\alpha$ near $45^\circ$
and $135^\circ$,
where the dependence of $\sin(2\alpha)$ on $\alpha$ is lowest.
In Fig.~\ref{onesigma}b, c, and d, we show the errors for the case where
$A_P = 0.2$ and we are able to put the specified constraints on the
amplitudes.  We see that in the case where there is a penguin amplitude,
and it is well understood, in general, the errors are smaller than in the
case of no penguin amplitude.

\begin{figure}[hp]
\epsfysize=6in
\centerline{
\epsffile{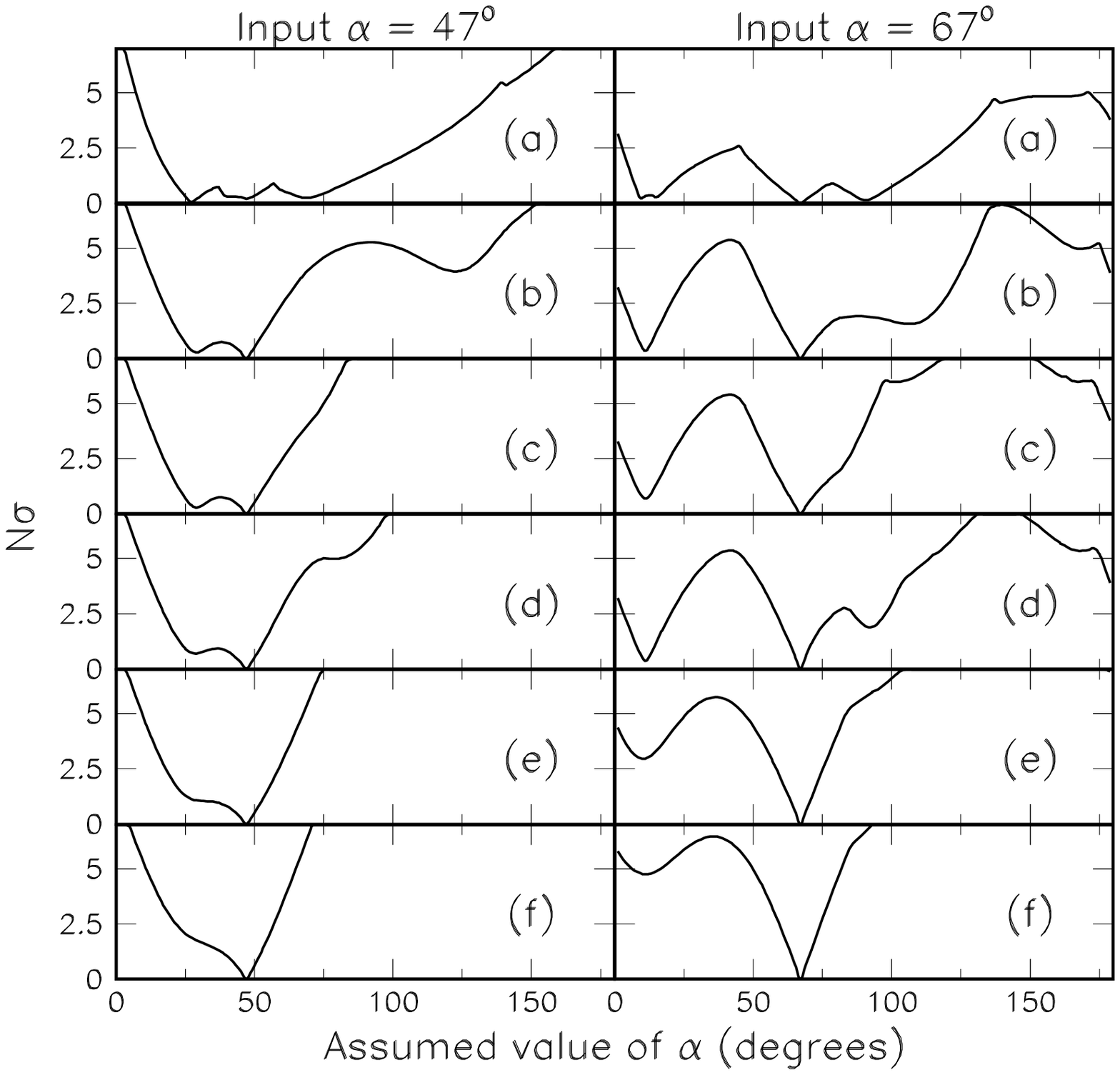}}
\caption{$N\sigma$ as a function of $\alpha$, for an input value of
$\alpha = 47^\circ$ on the left and $\alpha = 67^\circ$ on the
right.  The input value for $A_P/A_T$ is 0.2.
\hfil\break
For the following, we assume $\sigma(\avbr)/\avbr = 0.06$: \hfil\break
(a)  $\sigma(A_P)/A_P = 1.0$,
$\sigma(A_T)/A_T = 1.0$, no constraint on $\delta$. \hfil\break
(b)  $\sigma(A_P)/A_P = 1.0$,
$\sigma(A_T)/A_T = 0.03$, no constraint on $\delta$. \hfil\break
(c)  $\sigma(A_P)/A_P = 1.0$,
$\sigma(A_T)/A_T = 0.03$, $\sigma(\delta) = 20^\circ$. \hfil\break
(d)  $\sigma(A_P)/A_P = 0.1$,
$\sigma(A_T)/A_T = 0.03$, no constraint on $\delta$. \hfil\break
(e)  $\sigma(A_P)/A_P = 0.1$,
$\sigma(A_T)/A_T = 0.03$, $\sigma(\delta) = 20^\circ$. \hfil\break
For the following, we assume $\sigma(\avbr)/\avbr = 0.03$: \hfil\break
(f)  $\sigma(A_P)/A_P = 0.1$,
$\sigma(A_T)/A_T = 0.015$, $\sigma(\delta) = 20^\circ$
}
\label{nsigma}
\end{figure}

The plots in Fig.~\ref{onesigma} do not convey all the relevant information
on the constraints on $\alpha$.  The errors are highly non-Gaussian, and there
are multiple minima.  To gain further insight, we choose two particular
input values for $\alpha$, $47^\circ$ and $67^\circ$.  We then scan as a
function of the assumed value of $\alpha$ in the fit.  For each point in the
scan, we hold $\alpha$ fixed, and mimimize the $\chi^2$ with respect to all
the other parameters.  We then plot $\sqrt{\chi^2-\chi^2_{min}}$
as a function of $\alpha$,
interpreting $\sqrt{\chi^2-\chi^2_{min}}$ as the number of standard deviations
on $\alpha$.

We show the results in Fig.~\ref{ap05} for the case $A_P/A_T = 0.05$.
In Fig.~\ref{ap05}a, we assume
very little knowledge of the tree and penguin parameters.
We see that when the penguin contribution is small,
loose constraints are sufficient for the determination of $\alpha$.
However, even with the tight constraints of Fig.~\ref{ap05}e we are
unable to lift the discrete ambiguities.

We show the results in Fig.~\ref{nsigma} for the case $A_P/A_T = 0.2$.
In Fig.~\ref{nsigma}a, we assume
very little knowledge of the tree and penguin parameters.  As
qualitatively pointed out in Ref~\cite{penguins4},
in this case, we can rule out
only a small fraction of the available parameter space.  As we add
constraints in Fig.~\ref{nsigma}b, c, and d,  there are fewer minima
and more of the parameter space can be ruled out.  As shown in
Fig.~\ref{nsigma}e, it is not until we constrain all the parameters that
we are left with a single minimum.  Tightening the constraints in
Fig.~\ref{nsigma}f, we more convincingly rule out alternative minima and
improve the precision of the measurement of $\alpha$.

\begin{figure}[hp]
\epsfysize=6in
\centerline{
\epsffile{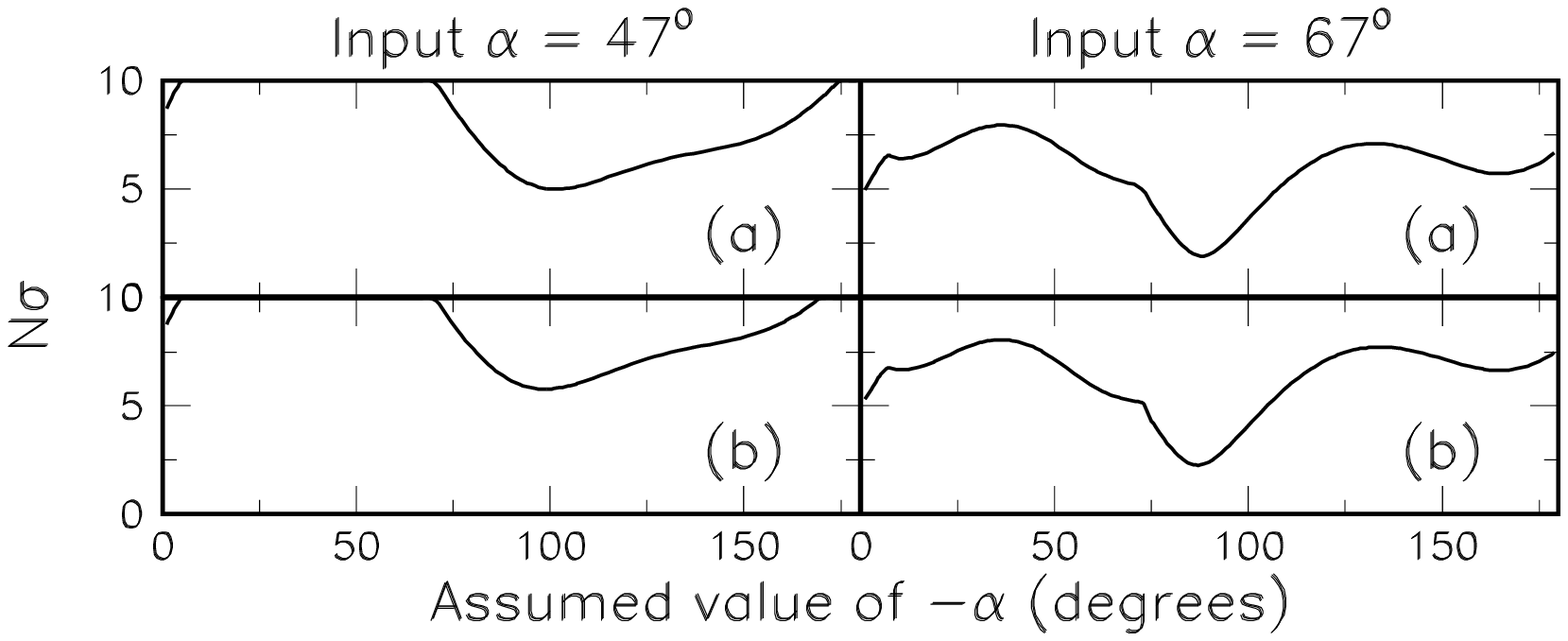}}
\vspace{-7cm}
\caption{$N\sigma$ as a function of $-\alpha$, for an input value of
$\alpha = 47^\circ$ on the left and $\alpha = 67^\circ$ on the
right.
For the following, we assume $\sigma(\avbr)/\avbr = 0.06$: \hfil\break
(a) $\sigma(A_P)/A_P = 0.1$,
$\sigma(A_T)/A_T = 0.03$, $\sigma(\delta) = 20^\circ$. \hfil\break
For the following, we assume $\sigma(\avbr)/\avbr = 0.03$:
(b)$\sigma(A_P)/A_P = 0.1$,
$\sigma(A_T)/A_T = 0.015$, $\sigma(\delta) = 20^\circ$.
}
\label{reverse}
\end{figure}

In Fig.~\ref{reverse}, we show similar plots as for Fig.~\ref{nsigma},
using the same input values for $\alpha$ but negative assumed values
for $\alpha$.  As discussed earlier, there is another solution only
for the discrete case $A=0$.  However, even for other values of $A$,
two solutions may be allowed within the uncertainties of the measurement.
Fig.~\ref{reverse} illustrates how well we can choose between the
solutions.  The separation is quite convincing for $\alpha=47^\circ$,
i.e. for large values of $A$, and
becomes more difficult for $\alpha = 67^\circ$, i.e. as $A$ gets smaller.

\begin{figure}[hp]
\epsfysize=3.5in
\centerline{
\epsffile{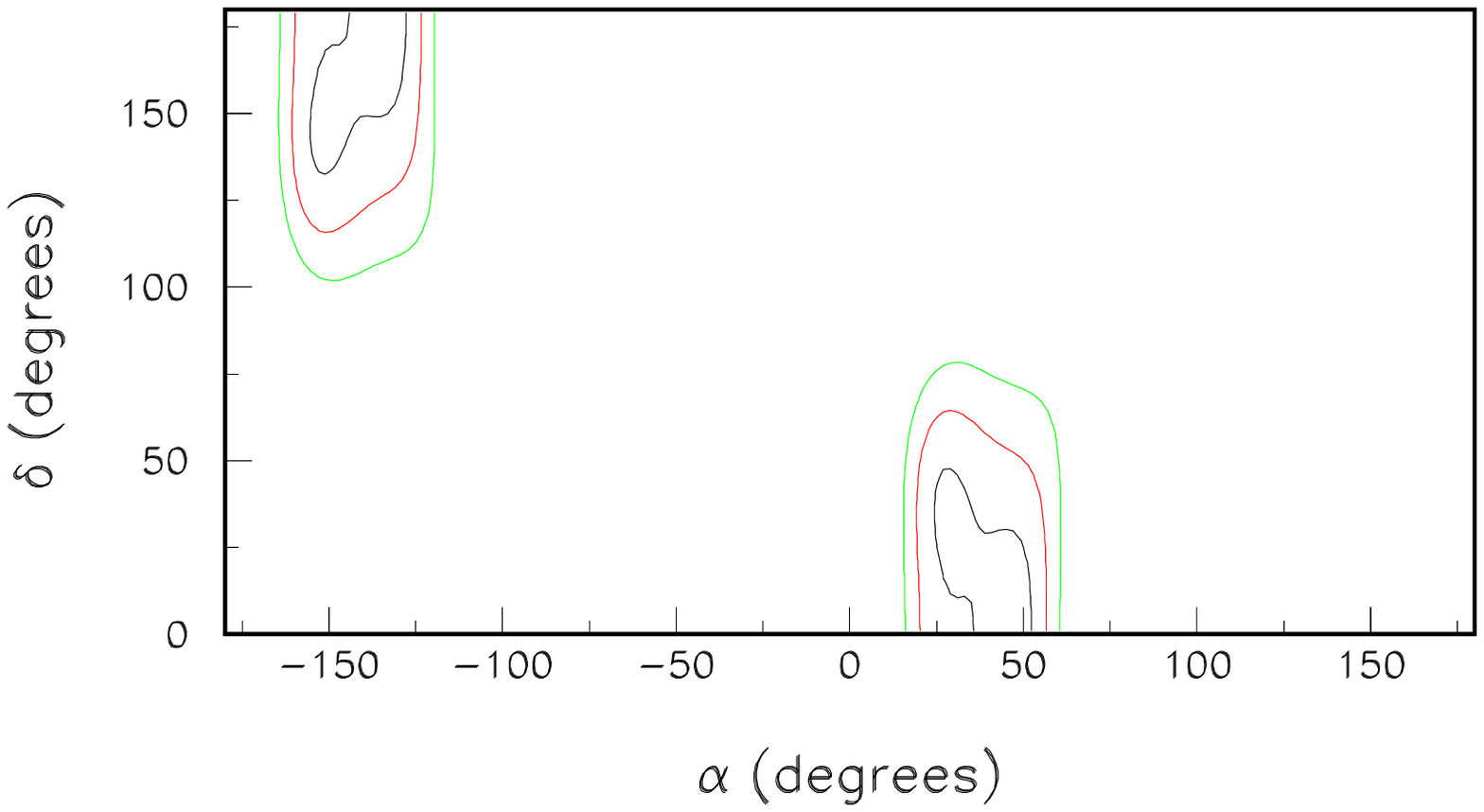}}
\vspace{-1.0cm}
\caption{One, two, and three $\sigma$ contours as a function of $\alpha$
and $\delta$, for input values of $\alpha = 47^\circ$, $\delta = 0^\circ$,
and $A_P/A_T$ = 0.2.}
\label{alpha2d_1}
\epsfysize=3.5in
\centerline{
\epsffile{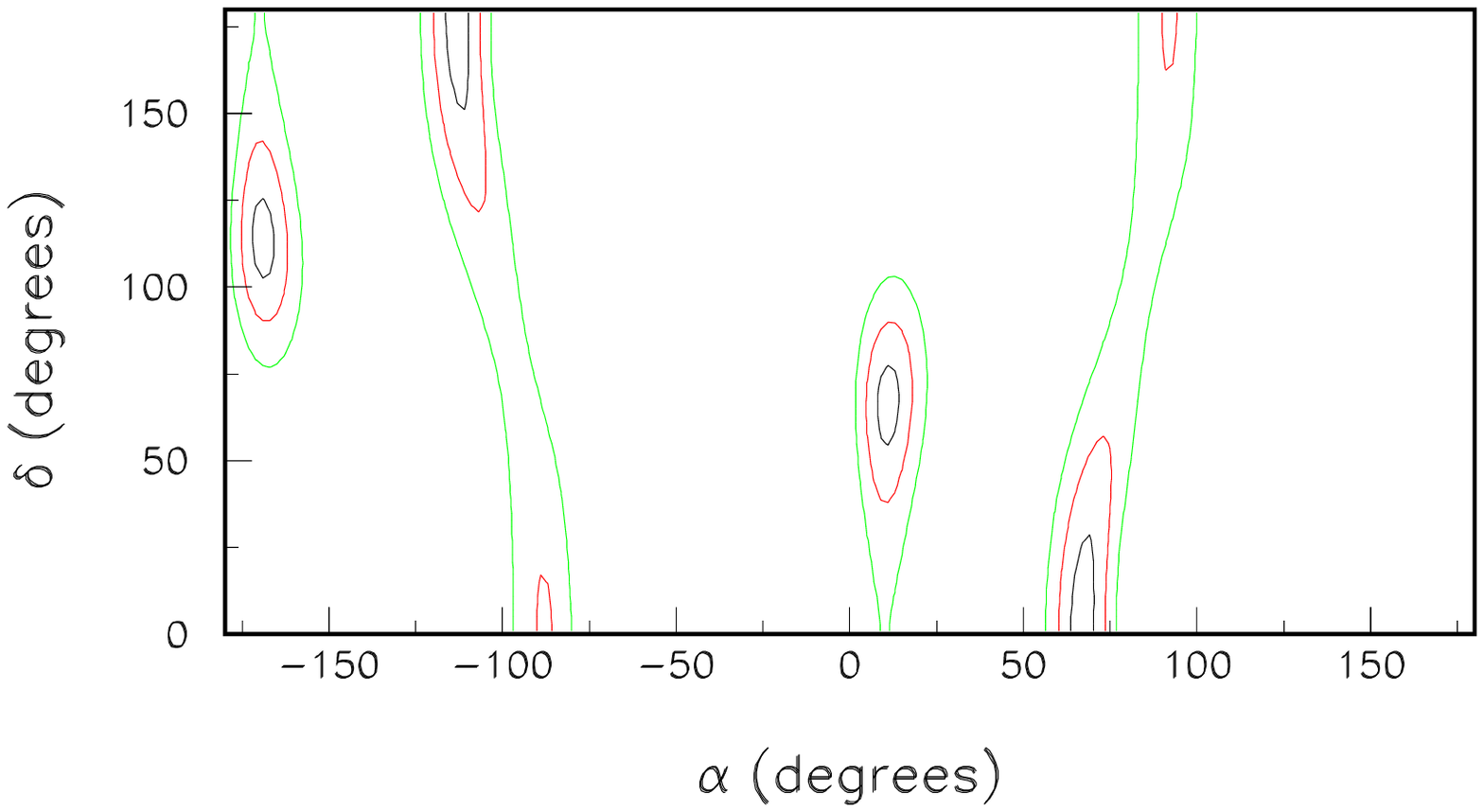}}
\vspace{-1.0cm}
\caption{One, two, and three $\sigma$ contours as a function of $\alpha$
and $\delta$, for input values of $\alpha = 67^\circ$, $\delta = 0^\circ$,
and $A_P/A_T$ = 0.2.}
\label{alpha2d_2}
\end{figure}

It is clear from the above plots that knowledge of the strong phase
difference $\delta$ decreases the error on $\alpha$ and in addition
can help distinguish between the discrete solutions on $\alpha$.
Figures~\ref{alpha2d_1} and~\ref{alpha2d_2}
are similar to Figure~\ref{nsigma}, except we scan as a function of
assumed values for both $\alpha$ and $\delta$, and plot
constant contours in $\sqrt{\chi^2-\chi^2_{min}}$.
In the case of $\alpha=47^\circ$ there are only two minima, and one
needs to know $\delta$ to better than $60^\circ$ in order to distinguish
between the two discrete solutions by $2\sigma$.
In the more complicated case of
$\alpha = 67^\circ$, there are multiple solutions even within
the first two quadrants for $\alpha$.  Here a more precise knowledge
of $\delta$ is required to unambiguously extract $\alpha$.

\begin{figure}[hp]
\epsfysize=6in
\centerline{
\epsffile{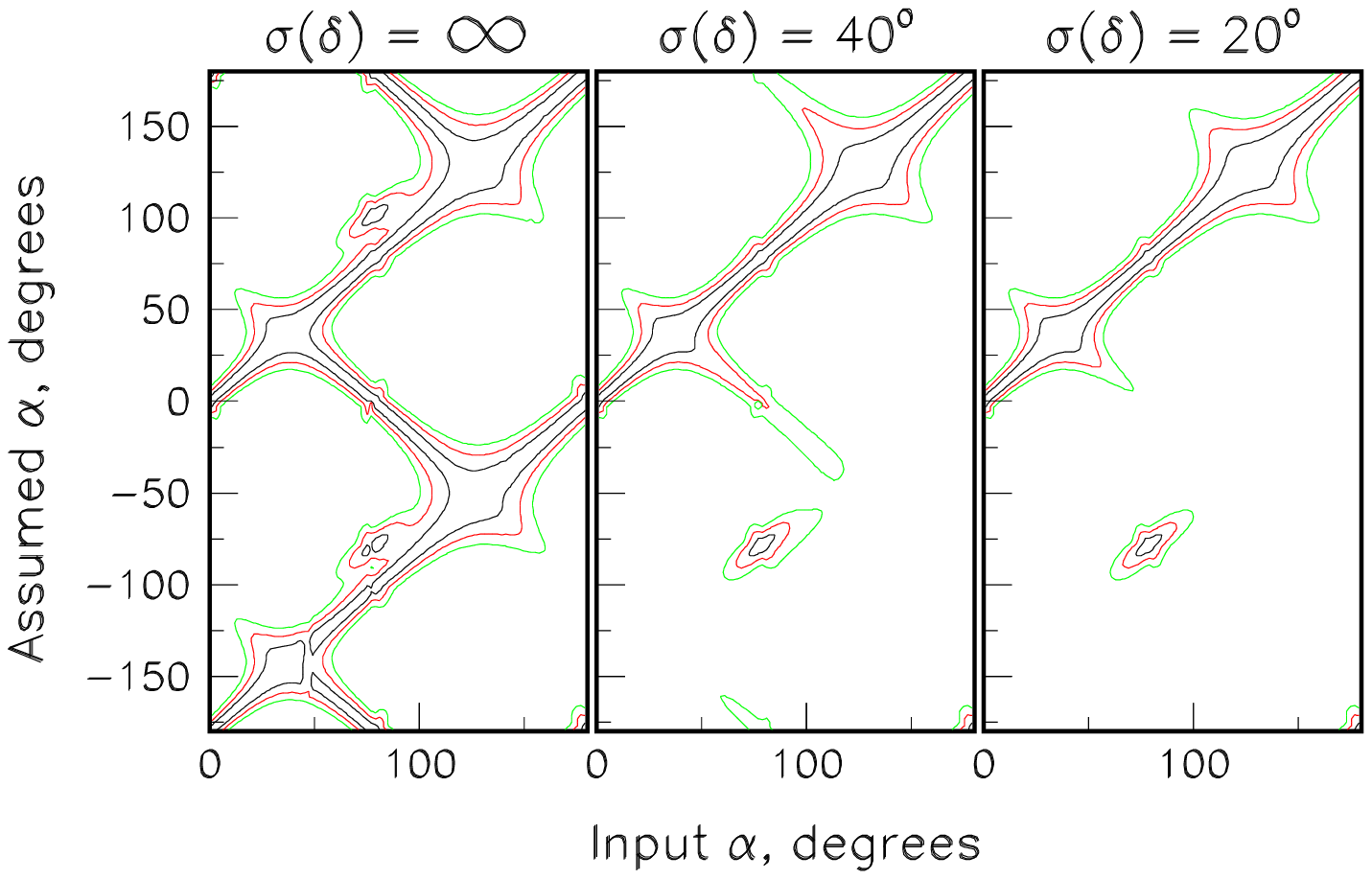}}
\vspace{-1.5cm}
\caption{One, two, and three $\sigma$ contours
as a function of input value of $\alpha$
and assumed value of $\alpha$, for various values
of $\sigma_\delta$.  Note that there always remains a discrete ambiguity
between $\alpha$ and $-\alpha$ where $A=0$.}
\label{alpha_v_alpha}
\end{figure}

The above two values of $\alpha$ of $47^\circ$ and $67^\circ$ lead to quite
different conclusions on the knowledge of $\delta$ required for a unique
determination of $\alpha$.
We have also investigated the effect of the uncertainty on $\delta$
on the extraction of $\alpha$, for all values of $\alpha$.
As an example,
Figure~\ref{alpha_v_alpha} shows constant contours of
$\sqrt{\chi^2-\chi^2_{min}}$ in the plane of input value of $\alpha$
and assumed value of $\alpha$, for three different constraints on
$\delta$.  We see that
the discrimination between discrete ambiguities improves with the
precision on $\delta$ until the uncertainty on $\delta$ reaches
a value of $20^\circ$.  We find that this result holds for all
values of $\alpha$.

\section{Conclusions}

We have presented the results of an error analysis on $\alpha$,
given a measurement of
the time-dependent asymmetry between
$B^0 \to \pi^+\pi^-$ and ${\bar B^0} \to \pi^+\pi^-$,
a measurement of the average branching ratio
for $B$ and $\bar{B}$ decays to $\pi^+\pi^-$,
and constraints on the magnitudes $A_T$ and $A_P$ and relative phase $\delta$
of the tree and penguin diagrams.  While there are an infinite number
of possible scenarios, our results set the scale for how well
these parameters need to be known.
We have considered scenarios in which we have an effective number
of 100 perfectly tagged signal events\cite{dilution}.

In the case where the penguin
contribution is small ($A_P / A_T \approx 0.05$), only crude
information on $A_P$ and $A_T$ is needed for the extraction of
$\alpha$.  However, we are left with four discrete ambiguities,
as in the case where there is no penguin contribution.

In the case where the penguin contribution
is larger ($A_P / A_T \approx 0.2$), more precise information
on $A_P$ and $A_T$ is needed.  If this precision can be achieved,
the uncertainty on $\alpha$ is in many cases smaller than for the
case of no penguin amplitude.  Furthermore, if it is possible to
place constraints on $\delta$, some or all of the discrete ambiguities
may be lifted.  A large penguin amplitude therefore presents an
opportunity for a much improved determination of $\alpha$.

In summary, we find that if the penguin
amplitudes are either small or well understood then it is
possible to determine $\alpha$ from the CP asymmetry in
$B^0 \to \pi^+\pi^-$ without resorting to the observation of
final states with neutral particles.
Thus, measurements would be feasible at hadron colliders as
well as $e^+ e^-$ colliders.  Furthermore, a large penguin
amplitude presents an opportunity for improved precision on $\alpha$
while lifting some or all of the discrete ambiguities.

\section{Acknowledgements}
This study was performed in the context of discussions on the physics
goals and detector upgrades for CDF in Run 2.  We thank our collaborators
for their kind advice and comments.  In addition, we thank
Isi Dunietz, Chris Hill, Jonathan Lewis, and Jon Rosner for useful
discussions.

\newpage

\appendix
\section{Likelihood formalism for the extraction of fitting errors}

\setcounter{equation}{0}

The analysis follows closely that in reference \cite{McDonald}.  For brevity,
we define the functions $n_\pm(t)$:
\begin{eqnarray}
n_\pm(A,\phi,t) = 1 \pm A \sin ( xt +\phi)
\end{eqnarray}
With this notation, and ignoring an irrelevant constant term,
the log-likelihood (see equation~\ref{likelihood}) is
\begin{eqnarray}
\label{logl}
\ln {\cal L}(A,\phi) = \sum_i \ln n_+(A,\phi,t_i) + \sum_j \ln n_-(A,\phi,t_j)
\end{eqnarray}
The second derivatives of the above function are the elements of the inverse
of the correlation matrix, $G = V^{-1}$.  For example,
\begin{eqnarray}
\label{errA}
G_{AA} = -\frac{\partial^2 \ln {\cal L}}{\partial A^2} =
{\sum_{i}  \frac{\sin^2(xt_i+\phi)}{[n_+(t_i)]^2}} +
{\sum_{j}  \frac{\sin^2(xt_j+\phi)}{[n_-(t_j)]^2}}
\end{eqnarray}
These sums can be approximated by integrals:
\begin{eqnarray}
\sum_k f_\pm(t_k) \approx \int_{T}^{\infty} f_\pm(t)N_\pm(t) dt
\end{eqnarray}
where the limits of integration assume that a lifetime cut is imposed
on the reconstructed mesons, and $N_\pm(t)$ are defined in
equation~\ref{eqn}.
Since in general $f_\pm(t) = g_\pm(t)/[n_\pm(t_i)]^2$,
equation~\ref{errA} (for the general error on variables $p$ and $q$) thus
becomes
\begin{eqnarray}
G_{pq} & = & \frac{N}{2} {\int_{T}^{\infty} e^{-t} \left [
\frac{g_+(t)}{n_+(t)} + \frac{g_-(t)}{n_-(t)} \right ]dt} \\
 & = & \frac{N}{2} {\int_{T}^{\infty} e^{-t}
	\left [ g_+(t)n_-(t)+g_-(t)n_+(t)\right ] dt}
\end{eqnarray}
where we have ignored terms which of order $A^2$:
\begin{eqnarray}
n_+(t)n_-(t) = 1 - A^2 \sin ^2 (xt +\phi) \approx 1.
\end{eqnarray}
We expect this to be a reasonable approximation even for extreme values
of $\sin(2\alpha)$, since in practice the observed asymmetry will be reduced
by a dilution factor $D \sim 0.5$ resulting from imperfect flavor tagging.
With some algebra, we obtain
\begin{eqnarray}
\label{errors}
G_{AA} & = & N {\int_{T}^{\infty}
		e^{-t} \sin ^2 (xt +\phi) dt} \\
G_{\phi\phi} & = & N A^2 {\int_{T}^{\infty}
		e^{-t} \cos ^2 (xt +\phi) dt} \\
G_{A\phi} & = & \frac{N}{2} A {\int_{T}^{\infty}
		e^{-t}  \sin 2(xt +\phi) dt}
\end{eqnarray}
We note that, as expected, these equations are invariant with respect to
the following transformation in the presence of dilutions:
\begin{eqnarray}
A & \rightarrow & DA \\
N & \rightarrow & \frac{N}{D^2}
\end{eqnarray}
For brevity, we introduce two new functions, $P_c(x,T)$ and $P_s(x,T)$ given by
\begin{eqnarray}
\label{Pfunctions}
P_c(x,T) = \frac{1}{1+x^2} \left [ \cos xT - x \sin xT \right ] \\
P_s(x,T) = \frac{1}{1+x^2} \left [ x \cos xT + \sin xT \right ]
\end{eqnarray}
and the end result is
\begin{eqnarray}
\label{eqnresult}
G_{AA} & = & \frac{N}{2}e^{-T}
		\left [ 1 - P_c(2x,t_o)\right ] \\
G_{\phi\phi} & = & A^2 \frac{N}{2}e^{-T}
		\left [ 1 + P_c(2x,t_o)\right ] \\
G_{A\phi} & = & \frac{N}{2} e^{-T} A P_s(2x,t_o)
\end{eqnarray}
where $t_o= T + {\phi \over x}$.

\newpage

\def\np#1#2#3 {Nucl. Phys. {\bf B#1} #2, (#3)}
\def\zpc#1#2#3 {Zeit. Phys. {\bf C#1} #2, (#3)}
\def\pl#1#2#3 {Phys. Lett. {\bf B#1} #2, (#3)}
\def\prl#1#2#3 {Phys. Rev. Letters {\bf #1} #2, (#3)}
\def\prd#1#2#3 {Phys. Rev. {\bf D#1} #2, (#3)}


\begin{thebibliography}{99}
\bibitem{dilution} For a sample of $S$ signal events, on
top of a background of $B$ events, the uncertainty on
the CP asymmetry, $\delta ( A_{CP})$, is given by
$\delta( A_{CP})^2 = (S+B)/\epsilon D^2 S^2$ where $\epsilon$ is
the efficiency of the flavor-tagging algorithm and $D$ is
the ``dilution'' of the algorithm, defined as
$D = (N_R-N_W)/(N_R+N_W)$ where $N_R, N_W$ are the number
of correct and incorrect tags respectively.  The effective number
of tagged signal events is thus $N_{tag} = \frac{S}{S+B}\epsilon D^2 S$.
\end{thebibliography}
\end{document}